\documentclass{article}

\usepackage[preprint]{preprint}

\usepackage[utf8]{inputenc} 
\usepackage[T1]{fontenc}    
\usepackage{hyperref}       
\usepackage{url}            
\usepackage{booktabs}       
\usepackage{amsfonts}       
\usepackage{nicefrac}       
\usepackage{microtype}      
\usepackage{bm}
\usepackage{wrapfig}

\usepackage[dvipsnames]{xcolor}

\usepackage{amssymb}
\usepackage{amsmath}
\usepackage[pdftex]{graphicx}

\title{
Spherical convolutions on molecular graphs for protein model quality assessment
}

\author{%
Ilia ~Igashov\thanks{
	These authors contributed equally.
}\\
  Moscow Institute of Physics and Technology\\
  Univ. Grenoble Alpes, Inria, CNRS,\\ Grenoble INP, LJK, 38000 Grenoble, France \\
  \texttt{igashov.is@phystech.edu} \\
  \And
  Nikita ~Pavlichenko\footnotemark[1]\\
  Moscow Institute of Physics and Technology\\
 Dolgoprudny, Moscow region, 141700,  Russia \\
  \texttt{pavlichenko.nv@phystech.edu} \\
  \And
  Sergei ~Grudinin \\
  Univ. Grenoble Alpes, Inria, CNRS, Grenoble INP, LJK, 38000 Grenoble, France \\
  \texttt{sergei.grudinin@inria.fr} \\
}

\begin{document}

\maketitle

\begin{abstract}

Processing information on 3D objects requires methods stable to rigid-body transformations, in particular rotations, of the input data. In image processing tasks, convolutional neural networks achieve this property using rotation-equivariant operations. However, contrary to images, graphs generally have irregular topology. This makes it challenging to define a rotation-equivariant convolution operation on these structures. 

In this work, we propose Spherical Graph Convolutional Network (S-GCN) that processes 3D models of proteins represented as molecular graphs. In a protein molecule, individual amino acids have common topological elements. This allows us to unambiguously associate each amino acid with a local coordinate system and construct rotation-equivariant spherical filters that operate on angular information between graph nodes. Within the framework of the protein model quality assessment problem, we demonstrate that the proposed spherical convolution method significantly improves the quality of model assessment compared to the standard message-passing approach. It is also comparable to state-of-the-art methods, as we demonstrate on Critical Assessment of Structure Prediction (CASP) benchmarks. The proposed technique operates only on geometric features of protein 3D models. This makes it universal and applicable to any other geometric-learning task where the graph structure allows constructing local coordinate systems.
We will make the method available at \url{https://team.inria.fr/nano-d/software/s-gcn/}.
\end{abstract}

\section{Introduction}

Prediction of protein three-dimensional (3D) structure is an important problem in structural biology and structural bioinformatics. 
Despite tremendous progress in this field \citep{greener2019deep, xu2019distance,senior2020improved,kryshtafovych2019critical}, 
particularly in light of the recent CASP14 results \citep{Callaway_2020},
the accuracy of the predicted structures tends to vary significantly depending on the availability of additional information, and the number of homologous structures and sequences in the databases \citep{abriata2019further,
senior2019protein,zheng2019deep,hou2019protein}. 
Therefore, estimation of reliability of the predicted models, and also the assessment of the local structural fragments, is crucial for the practical application of these predictions.

The problem of {\em protein model quality assessment} (MQA) has been recognized by the protein structure modeling community and became one of the subchallenges of CASP, the Critical Assessment of protein Structure Prediction community-wide challenge \citep{cheng2019estimation,won2019assessment}. 
All of the state-of-the-art MQA methods use, to a certain extent, supervised or unsupervised machine learning. 
Initially, statistical potentials \citep{olechnovivc2017voromqa}, shallow neural networks \citep{Wallner2003}, regression methods, and support vector machines \citep{Ray2012,Uziela2016} were widely used.
More recently, this problem has also got attention from the machine-learning community. 
This triggered the development of more advanced approaches, such as deep learning-based techniques \citep{Derevyanko2018,pages2019protein,Conover2019,hiranuma2020improved,jing2020learning,eismann2020protein} 
and graph convolutional networks (GCN) \citep{Graph-QA,sanyal2020proteingcn,igashov2020vorocnn}.
The latter methods operate on a {\em molecular graph} representation of protein {\em models}.

In this work, we propose
to capture the 3D structure of a molecular graph using convolution operation based on spherical harmonics.
The main idea of our approach is to learn spatial filters in a reference orientation of each graph node.
Indeed, proteins are chained molecules, with a repeated topology of the backbone.
Thus, using local coordinate frames constructed on the protein's backbone, we can build {\em rotational-equivariant} spherical filters.
We then incorporate these filters into a message-passing framework
and design  a new method called {\em Spherical  Graph Convolutional Network (S-GCN)},
which significantly outperforms the classical GCN architecture.

Most of the protein MQA methods operate on the atom-level representation of a protein molecule \citep{Uziela2016,olechnovivc2017voromqa,karasikov2019smooth,pages2019protein,igashov2020vorocnn}.
At the same time, the state-of-the-art methods use various types of features that  often include information about the evolution of molecules  or other biological characteristics.
On the contrary, S-GCN works with the residue-level protein representation, which significantly reduces computational costs and the number of parameters.
Also, our method processes only geometric information, i.e. the input feature vector of each amino acid contains only three geometric features and a one-hot vector representing the type of the amino acid.
This work demonstrates that the state-of-the-art quality of the protein model assessment task 
can be achieved using only geometric properties of protein models without any chemo-physical prior information, 
which often requires additional expensive computations.

The main results of our work can be summarized as follows:
\begin{itemize}
	\item We propose a new message-passing method based on trainable rotational-equivariant spherical filters.
	\item The proposed method significantly improves the quality of model assessment as compared to a classical GCN approach, applied to the same input configuration.
	\item Despite the residue-level representation and only geometric input features, the results of the proposed method are comparable to the state of the art.
\end{itemize}

\section {Related work}

\textbf{Structural bioinformatics.}
Quality assessment of protein models is a classical problem in protein structure prediction community. 
There have been multiple approaches developed over last 30 years.
These include physics-based techniques \citep{Randall2008,Faraggi:2014aa},
statistical and unsupervised methods, such as  DFIRE \citep{Zhou2002}, DOPE \citep{Shen2006}, GOAP \citep{Zhou2011}, RWplus \citep{Zhang:2010aa}, ORDER\_AVE \citep{Liu2014}, VoroMQA \citep{olechnovivc2014voronota} and more,
classical ML-approaches ModelEvaluator \citep{Wang2009}, ProQ2 \citep{Ray2012}, Wang\_SVM~\citep{Liu2016}, Qprob~\citep{Cao:2016aa}, SBROD \citep{karasikov2019smooth}, a learning-to-rank technique \citep{Jing2016}, 
deep learning methods \citep{Derevyanko2018,pages2019protein,Conover2019,sato2019protein,Jing:2020qy,hiranuma2020improved}, neural \citep{Wallner2003},
and graph neural networks  \citep{Graph-QA,sanyal2020proteingcn,igashov2020vorocnn}.

\textbf{Graph neural networks (GNNs) for molecular graphs.}
	In the last years, various GNNs were proposed to address the problem of learning on molecular graphs. 
	Starting with the message-passing paradigm in the molecular graph domain \citep{gilmer2017neural}, 
	further multiple approaches elaborated on this idea \citep{schutt2017schnet,thomas2018tensor,chen2019graph,nachmani2020molecule,sun2020graph,klicpera2020directional}. 
	All of them were designed to operate on small molecules.
	For example, in QM9 \citep{rupp,blum}, a popular benchmark that is used to evaluate these methods, molecules consist of up to 23 atoms. 
	On the contrary, a protein molecule can contain thousands of atoms, and this fact requires different approaches that take into account the size of the data.
	Recently, GNNs have also been started to be applied to protein graphs for solving various problems such as protein design \citep{ingraham2019generative}, protein docking \citep{fout2017protein,cao2019energy}, classification \citep{weiler18073d,zamora2019structural}, and quality assessment \citep{sanyal2020proteingcn,Graph-QA,igashov2020vorocnn}.

\textbf{Equivariance.}
Processing information in 3D must be stable against rigid-body transformations of the input data. 
This stability can be achieved using {\em equivariant operations},
which is a very active research topic, especially regarding rotational equivariance.
For example, rotation-equivariant CNNs were proposed for  spherical images using correlations on a sphere \citep{cohen2018spherical} and then extended to fully Fourier-space architectures \citep{kondor2018clebsch,anderson2019cormorant}.
Similar architectures can be constructed for rigid-body motions using tensor field rotation- and translation-equivariant networks \citep{thomas2018tensor,weiler18073d}.
Spherical harmonics kernels have also been applied to point-cloud data \citep{poulenard2019effective}.
Alternatively, for some types of volumetric data, rotation-equivariant representation can be constructed with oriented local coordinate frames \citep{pages2019protein}.
The same idea can be applied to the protein graph representation, where the spatial relation between local frames can be encoded using spatial edge features  \citep{ingraham2019generative} or additional edge descriptors \citep{sanyal2020proteingcn}.
For general molecular graphs, the problem is more difficult.
Still, there has been significant progress using, e.g. the message-passing formalism with messages containing radial and directional information about neighboring graph nodes  \citep{klicpera2020directional}.

\section{Proposed Method}

\subsection{Protein graph}

\begin{figure}[h!]
	\centering
	\includegraphics[width=1\textwidth]{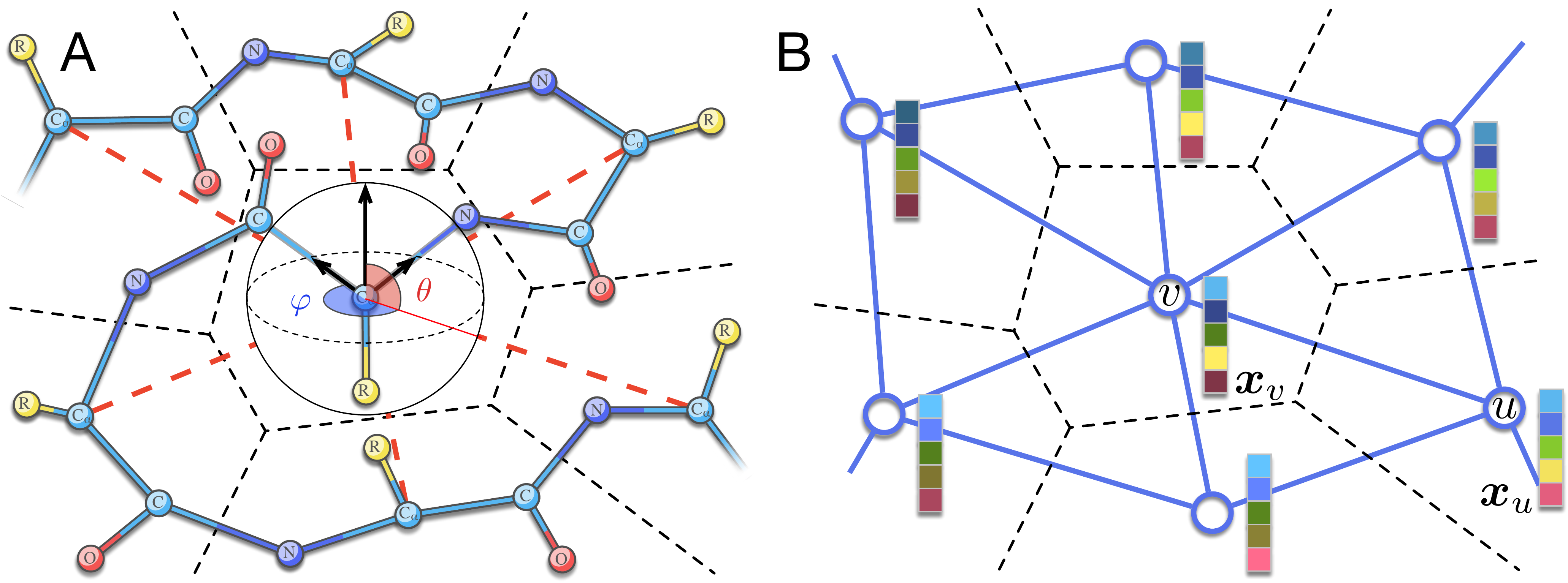}
	\caption{\label{fig:scheme} 
	Schematic representation of a molecular graph.
	(A) 3D protein structure is partitioned into Voronoi cells, shown with the dashed lines. 
	The central amino acid has the associated coordinate system, which is built according to the topology of its backbone (atoms $C$, $C_{\alpha}$, $N$) with the center at the position of  the $C_{\alpha}$ atom.
	$R$ symbols denote amino acid residues.
	The spherical angles  $\varphi$ and $\theta$ of the neighboring residues are computed with respect to the local coordinate system of the central residue.
	(B) Graph corresponding to the Voronoi tessellation,
	$v$ is the central node, $u$ is its neighbor,  $\boldsymbol{x}_v$ and  $\boldsymbol{x}_u$ are the corresponding feature vectors, which are also shown with colored boxes.
	}
\end{figure}

A protein molecule is a chain of amino acids, or {\em residues}, folded in a 3D space.
We construct a graph $\mathcal{G}$ of the protein molecule by splitting the surrounding space into cells using the Voronoi tessellation method Voronota \citep{olechnovivc2014voronota}.
Nodes of the resulting graph correspond to the protein residues and edges are associated with the pairs of residues whose Voronoi cells have a non-zero contact surface. Figure \ref{fig:scheme} schematically shows the graph construction.

Each node $v$ of the graph $\mathcal{G}$ contains a feature vector $\boldsymbol{x}_v$ associated with the corresponding protein residue.
These features include one of $20$ amino-acid types encoded with the one-hot representation, the solvent-accessible surface area for each residue, the volume of residue’s Voronoi cell, and the “buriedness” of the residue, which is a topological distance in the graph $\mathcal{G}$ to the nearest solvent-accessible node.
We represent the whole set of nodes as a feature matrix $\textbf{X}\in\mathbb{R}^{N\times{d}}$ where $N$ is a number of residues and $d=23$ is the size of the feature vector.

To describe the edges of the graph $\mathcal{G}$, we will use the following notations. 
Let $\boldsymbol{A}\in{\mathbb{R}^{N\times{N}}}$ be the symmetric binary adjacency matrix of the graph. 
For any pair of residues $v$ and $u$, the two corresponding entries of the matrix $\boldsymbol{A}$ equal $1$ if $u$ and $v$ have an edge, and zero otherwise.
In our settings, the graph $\mathcal{G}$ does not have {\em self-loops}, 
hence
the main diagonal elements of the matrix $\boldsymbol{A}$ are zeros.
In order to refer to neighbors of a node $v$ in the graph $\mathcal{G}$, 
i.e. those nodes that have a common edge with $v$,
we will use notation $\mathcal{N}(v)$.

\subsection{Spherical harmonics}

Let us consider a complex square-integrable function  $f(\theta, \varphi)$ defined on a unit sphere $S_1$.
This function can be expanded in a polynomial basis using {\em spherical harmonics} as the  basis functions,

\begin{equation}\label{expansion}
	f(\theta, \varphi) =\sum_{l = 0}^{\infty} \sum_{m=-l}^{l} w_l^m Y_l^m(\theta, \varphi),
\end{equation}

where $w_l^m$ are the expansion coefficients, 
and $Y_l^m(\theta, \varphi)$ are the spherical harmonics \citep{hobson1955theory},

\begin{equation}
	Y_l^m(\theta, \varphi) = \sqrt{\dfrac{(2l + 1)}{4\pi}\dfrac{(l-m)!}{(l+m)!}}P_l^m(\cos \theta)e^{im\varphi}.
\end{equation}

Here, $P_l^{m}(\cos \theta)$ are the associated Legendre polynomials \citep{hobson1955theory}.
We should also note that a real function on a unit sphere 
can be decomposed in a polynomial basis more compactly using  {\em real spherical harmonics} as the basis functions.
Below we will be using this real basis, which is specified in Supplementary Materials.

\subsection{Local coordinate system}

The protein backbone
consists of atom repetitions $C$, $C_{\alpha}$, $N$, $O$. 
This allows us to unambiguously associate each residue with a local coordinate system.
Indeed, for each residue we can define the normalized $C_{\alpha}$--$N$ vector as the $x$-axis, the unit vector lying in the $C$--$C_{\alpha}$--$N$ plane,  orthogonal to $x$, and having positive dot product with  $C_{\alpha}$--$C$ as the $y$-axis, and the vector product of $x$ with $y$ as the $z$-axis.
Then, given a node $v$, we can associate each neighbor $u\in\mathcal{N}(v)$ with a pair of {\em spherical angles} $\Omega_v^u = (\theta_v^u, \varphi_v^u)$.
They specify the angular position of the projection of the node $u$ onto a unit sphere in the local coordinate system of $v$.
An example of a local coordinates system is schematically shown in Figure \ref{fig:scheme}A.
Now, having an unambiguous orientation for each node in the graph, we can construct a {\em rotation-equivariant} convolution operation.

\subsection{Spherical convolution}

We can approximate the expansion \eqref{expansion} of the function $f(\theta, \varphi)$ by cutting the series at the maximum expansion order $L$,
\begin{equation}\label{expansion}
 	f(\theta, \varphi) \approx \hat{f}(\theta, \varphi) = \sum_{l=0}^{L}\sum_{m=-l}^l w_l^m Y_l^m(\theta, \varphi).
\end{equation}
The same approximation can be obtained for a matrix function $\boldsymbol{F}:S_1\to\mathbb{R}^{d_1\times{d_2}}$, $d_1,d_2\in\mathbb{N}$,
\begin{equation}\label{matrix_expansion}
\boldsymbol{F}(\theta, \varphi) \approx \hat{\boldsymbol{F}}(\theta, \varphi) = \sum_{l=0}^{L}\sum_{m=-l}^l \boldsymbol{W}_l^m Y_l^m(\theta, \varphi),
\end{equation}
where matrices $\boldsymbol{W}_l^m$ denote expansion coefficients of the function $\boldsymbol{F}$  in the   $Y_l^m$ basis.
Finally,
we can introduce the spherical convolution operation for the vertex $v$ in the following way,

\begin{equation}\label{convolution}
	\boldsymbol{F}\circ v = \sum_{u \in \mathcal{N}(v)} \hat{\boldsymbol{F}}(\theta_v^u, \varphi_v^u)\boldsymbol{x}_v.
\end{equation}

Considering matrices $\boldsymbol{W}_l^m$ to be optimized parameters, we will thus learn a spherical filter.
We should specifically emphasize that matrices  $\textbf{W}_l^m$ are rotation-equivariant by construction.

\subsection{Neural network}
The distinctive feature of convolutional networks built on spatial graphs is the way the graph nodes exchange information  by passing messages to each other.
On each layer of the network, nodes' feature vectors are combined and updated using the information from the neighboring nodes \citep{Scarselli2009,Kipf2016}.
In our implementation, for the information exchange, we use the proposed spherical convolution operation  \eqref{convolution}.

Let $\boldsymbol{A}_{\Omega}\in{\mathbb{R}^{N\times{N}}}$ be a matrix of local angular coordinates for each node's neighbor in the adjacency matrix  $\boldsymbol{A}$.
This means, for any pair of graph nodes $v$ and $u$ connected with an edge, the corresponding entry of matrix $\boldsymbol{A}_{\Omega}$ is a pair $\Omega_v^u=(\theta_v^u, \varphi_v^u)$ of angular coordinates of $u$ with respect to the local coordinate system of $v$.
We also denote
$Y_l^m(\boldsymbol{A}_{\Omega})\in{\mathbb{R}^{N\times{N}}}$
as a result of the elementwise application of the spherical harmonics $Y_l^m$ to the matrix $\boldsymbol{A}_{\Omega}$.
	We should note that the main diagonal elements of matrices $\boldsymbol{A}_{\Omega}$ and $Y_l^m(\boldsymbol{A}_{\Omega})$ are zeros,
	and, opposed to the adjacency matrix $\boldsymbol{A}$, matrices$\boldsymbol{A}_{\Omega}$ and $Y_l^m(\boldsymbol{A}_{\Omega})$ are not symmetric.
Then, the $k$th layer of the spherical graph convolutional network can be expressed as follows,

\begin{equation}\label{conv_layer}
  	\boldsymbol{H}^{k}=\sigma\left(\sum_{l,m}^{L}Y_l^m(\boldsymbol{A}_\Omega)\boldsymbol{H}^{k-1}\boldsymbol{W}_l^m + \boldsymbol{H}^{k-1}\boldsymbol{W} + \boldsymbol{b}\right),
\end{equation}
%
	where $\boldsymbol{H}^{k-1}\in{\mathbb{R}^{N\times{d_{k-1}}}}$ and $\boldsymbol{H}^{k}\in{\mathbb{R}^{N\times{d_{k}}}}$ are nodes' feature matrices before and after applying the layer, $\boldsymbol{H}^{0}\equiv\boldsymbol{X}$ is the input feature matrix, $\boldsymbol{W}_l^m\in{\mathbb{R}^{d_{k-1}\times{d_{k}}}}$ and $\boldsymbol{W}\in{\mathbb{R}^{d_{k-1}\times{d_{k}}}}$ are trainable parameters, $\boldsymbol{b}\in{\mathbb{R}^{d_k}}$ is a trainable bias vector, and $\sigma$ is a nonlinear activation function.
If we let the maximum expansion order $L=0$,
we can see that the operation \eqref{conv_layer} reduces to a standard message-passing form,

\begin{equation}\label{mp_layer}
	\boldsymbol{H}^{k} = \sigma\left(\boldsymbol{A}\boldsymbol{H}^{k-1}\boldsymbol{W}_0^0 + \boldsymbol{H}^{k-1}\boldsymbol{W} + \boldsymbol{b}\right),
\end{equation}
where for each node $v$, the first term transforms features from $v$'s neighbors $\mathcal{N}(v)$, 
the second term transforms and aggregates features from $v$ itself,
and the last terms provides a bias for the activation function $\sigma$.

In Supplementary Materials, we also provide a modification of the proposed spherical convolution layer \eqref{conv_layer}, which explicitly uses information about contact surface areas between Voronoi cells, and discuss the corresponding network architecture.

\section{Experiment}
The main purpose of the model quality assessment task is to evaluate the deviation between a generated model of a protein molecule and its native, or {\em target}, structure.
If the target structure is known, the quality of the model can be calculated by computing one of specifically designed metrics, e.g., CAD-score \citep{Olechnovic2012}, lDDT \citep{mariani2013lddt}, or GDT-TS \citep{zemla1999processing}.
Most often, however, experimental protein structures are unknown, and thus there is a need for protein structure prediction and model quality assessment.
In this section, we report the results of the protein model quality assessment task obtained by our spherical architectures and the GCN baseline.
We trained all of our networks using {\em local} per-residue CAD-scores as the ground truth.
They have been shown to be a more informative and stable metric compared to other MQA measures with respect to local structural perturbations of a protein molecule \citep{olechnovivc2019comparative}.
To  retrieve the {\em global} per-model scores, we averaged the predicted local scores.

We provide results of S-GCN along with several state-of-the-art MQA methods that are described in detail below. 
We would also like to emphasize that for several reasons, in this work, we do not compare S-GCN with recent GNNs designed for small molecules. 
First of all, we attempted to train tensor field networks \citep{thomas2018tensor} and DimeNet \citep{klicpera2020directional} on protein molecules but did not get any adequate results. 
Secondly, we are unable to test S-GCN on established ML benchmarks such as QM9 \citep{rupp,blum} due to the specificity of the graph representation in our method. 

\subsection{Datasets}

For our experiments, we collected  data from the Critical Assessment of Structure Prediction (CASP) benchmarks  \citep{moult1995large,kryshtafovych2019critical}.
They contain experimentally obtained native protein structures and the corresponding 3D models predicted by the CASP challenge participants.

For training, we used data from CASP[8-11] stage2 submissions.
For each target, we additionally generated 50 near-native models \citep{hoffmann2017nolb} in order to enrich the training dataset with high-quality examples.
For validation and selection of hyperparameters, we used data from CASP12 stage2 submissions.
All models that we used for training and validation were initially filtered and preprocessed.
More precisely, we excluded targets that had only models with low CAD-scores
and preprocessed all models by removing residues that were not present in the target structure.
More details and the list of all targets are available in Supplementary Materials.
In total, we had 333 target structures and 73,418 models from CASP[8-11] for training and 39 targets and 5,411 models from CASP12 for validation.
Finally, to test our architectures, we used unrefined data from CASP13 (73 target structures, 10,882 models) and unrefined data from CASP12 (38 target structures,  5471 models).

For each model, we precomputed
matrices $Y_l^m(A_{\Omega})$ up to the $10$th expansion order.
These matrices were the most space-consuming part of our dataset, as a spherical harmonic expansion of order $L$ requires the storage of $L^2$ coefficients for each pair of adjacent nodes in a graph.

\subsection{Metrics}
For the evaluation of the methods, we chose z-scores, MSE, determination coefficient $R^2$, Pearson, and Spearman correlations, as it is described in more detail below.
We used global CAD-scores as the ground truth for the assessment{
	\footnote{
			Although we mainly focus on experiments with CAD-score as the ground truth, we also evaluated the quality of our method on the same data with lDDT and GDT-TS as the ground truth. 
			The results are available in Supplementary Materials.
	}
}.
We computed z-scores for the top-predicted protein models for each target and then averaged them over all targets, as explained in more detail in Supplementary Materials.
For the MSE, $R^2$, and correlations, we used two different ways of calculation: {\em per-target} and {\em global}.
In the per-target approach, we computed the metrics separately within each protein target and then averaged results over all targets (we averaged correlations using the Fisher transformation \citep{fisher1915frequency}).
In the global approach, we stacked scores of all protein models into one vector and calculated the metrics on this vector.
For each metric, we also computed bootstrapped means and confidence intervals.
For the global metrics, a bootstrapped sample is chosen from the whole set of models.
For the per-target metrics, a bootstrapped sample is a sample of targets and their models, respectively.
These results are available in Supplementary Materials.

\begin{table}[h]
	\centering
	\caption{
		Architectures of our baseline and spherical graph convolutional networks.
		SCL and GCL are the spherical and graph convolution layers, correspondingly. FC is a fully-connected layer with ELU activation. 
		The parameters in the parentheses are the sizes of the input and the output feature vectors, correspondingly. 
		BN is the batch normalization layer.}
	\resizebox{\textwidth}{!}{\begin{tabular}{ll}
			\toprule
			Network&Architecture\\
			\midrule
			Baseline&\emph{\textbf{Encoder:}} FC(23, 32) $\to$ Dropout $\to$ FC(32, 64) $\to$ Dropout $\to$ FC(64, 128) $\to$ Dropout $\to$ \\
			& \emph{\textbf{Message-passing:}} GCL(128, 113) $\to$ Dropout $\to$ GCL(113, 98) $\to$ Dropout $\to$ \\
			&GCL(98, 83) $\to$ Dropout $\to$ GCL(83, 68) $\to$ Dropout $\to$ GCL(68, 53) $\to$ Dropout $\to$ \\
			&GCL(53, 38) $\to$ Dropout $\to$ GCL(38, 23) $\to$ Dropout $\to$GCL(23, 8) $\to$ Dropout \\
			&\emph{\textbf{Scorer:}} FC(8, 16) $\to$ Dropout $\to$ FC(16, 32) $\to$ Dropout $\to$ FC(32, 64) $\to$ Dropout $\to$ \\
			&FC(64, 32) $\to$ Dropout $\to$ FC(32, 16) $\to$ Dropout $\to$ FC(16, 1) $\to$ Sigmoid
			\\
			\midrule
			S-GCN&SCL(23, 20) $\to$ Dropout $\to$ SCL(20, 16) $\to$ BN $\to$ Dropout $\to$ SCL(16, 8) $\to$ Dropout $\to$ \\
			&SCL(8, 4) $\to$ BN $\to$ Dropout $\to$ SCL(4, 1) $\to$ Sigmoid\\
			\midrule
			S-GCN$_{\text{s}}$&\emph{\textbf{Spherical part:}} SCL(23, 20) $\to$ Dropout $\to$ SCL(20, 16) $\to$ BN $\to$ Dropout $\to$ \\
						 &SCL(16, 14) $\to$ Dropout $\to$ SCL(14, 12) $\to$ BN $\to$ Dropout $\to$ SCL(12, 8)$\to$ \\
						 & \emph{\textbf{Scorer:}} FC(8, 128) $\to$ D $\to$ FC(128, 64) $\to$ D $\to$ FC(64, 1) $\to$ Sigmoid\\
			\bottomrule
		\end{tabular}
		\label{tab:arc}
	}
\end{table}

\subsection{Baseline architecture}
For the baseline, we built a standard graph neural network based on the message-passing operation described in eq. \eqref{mp_layer}.
The structure of the proposed architecture can be split into three main parts.
The \emph{encoder} is a set of fully-connected layers that
transform the residues' features into a high-dimensional space. 
The \emph{message-passing part} is a set of graph convolution layers \eqref{mp_layer} that
capture the structure of a  protein graph  and work as a feature extractor.
Finally, the \emph{scorer} is a set of fully-connected layers
with a sigmoid at the end.
They form a multilayer perceptron and use the obtained features to predict the scores of each protein residue.
As a result, we obtain
three main design parameters -- the number of encoder, message-passing, and scoring layers.
We performed a grid search on the values of these parameters
and found out that the optimal architecture had 3 encoder
layers, 8 message passing layers, and 3 scoring layers. 
For each layer, we used the ELU activation function and the dropout rate
set to $0.3$, as we detected it to be optimal.
In total, our baseline network contains 339,053 trainable parameters.
Table \ref{tab:arc} briefly lists the final architecture.

\textbf{Training.} We trained this network on CASP[8-11] datasets for 40 iterations. 
We tuned hyperparameters on CASP12 (preprocessed) dataset.
For training, we used the Adam optimizer \citep{Kingma2014} and the Mean Squared Error (MSE) of local scores as the loss function.
On each iteration, we
trained the network in $4$ parallel processes feeding $512$ models to each process.
One training iteration took $\approx$21 min. on  Intel\textregistered Xeon\textregistered CPU E5-2630 v4 @ 2.20GHz, and $\approx$1 min. 10 sec. on NVIDIA GTX 1080 GPU.

\textbf{Hyperparameters.} The learning rate was $0.001$, the batch size $1$, the dropout rate $0.3$, and the $L_2$-regularization coefficient of $10^{-5}$.

\subsection{Spherical graph convolutional network architectures}

While constructing
a spherical graph convolutional network, we considered multiple expansion orders in the range from 3 to 10 and finally chose orders of $5$ and $10$.
We also experimented with the number of layers, the batch normalization layers \citep{ioffe2015batch} and the batch size, dropouts, the regularization parameters, and the output dimensionality of the spherical convolution layers.
The details of these experiments are available in Supplementary Materials.

Finally, we settled upon two architectures.
The first architecture, S-GCN, represents a sequence of spherical convolution layers \eqref{conv_layer} combined with dropout and batch normalization layers.
In the second architecture, S-GCN$_{\text{s}}$, we added three fully-connected layers to the end of the network following the idea used in the baseline.
S-GCN with the expansion order of $5$ contains $24,675$ trainable parameters, and S-GCN with the order of $10$ contains $95,475$ trainable parameters.
Respectively, S-GCN$_{\text{s}}$ with the order of $5$ contains $42,625$ trainable parameters, and S-GCN$_{\text{s}}$ with the order of $10$  contains $137,725$ trainable parameters.
Table \ref{tab:arc} briefly describes these configurations.

\textbf{Training.} For training, we used the Adam optimizer and the Mean Squared Error of local scores as the loss function.
We trained the networks on the shuffled data and split the whole training process into equal iterations.
Within each iteration, we trained each network in $4$ parallel processes feeding $2048$ models to each of them.
We stored and processed the adjacency matrices in a sparse format.
One training iteration takes on average  10 to 20 minutes, depending on the order of expansion on 
Intel\textregistered Xeon\textregistered CPU E5-2650 v2 @ 2.60GHz, and $\approx$12 minutes  on  NVIDIA GTX 1080 GPU for the 5th order S-GCN.
We trained 5th-order networks
for 40 iterations and 10th-order networks for 60 iterations.

\textbf{Hyperparameters.} 
The learning rate was set to $0.001$, the batch size $64$, the dropout rate $0.2$ and we used $L_2$-regularization with the coefficient of $0.003$ for the network of order 5.
We used the dropout rate of $0.1$ and $L_2$-regularization with the coefficient of $0.001$ for the network of order 10.

\begin{table}[t!]
	\centering
	\caption{
	Comparison of S-GCN and S-GCN$_{\text{s}}$ with the baseline network and the state-of-the-art MQA methods on the unrefined CASP12 stage2 dataset. 
	Parameters in parentheses correspond to the order of the spherical harmonic expansion.}
	\resizebox{\textwidth}{!}{\begin{tabular}{@{}llcllllcllll@{}}
			\toprule
			& 			            & & \multicolumn{4}{c}{Global metrics} & & \multicolumn{4}{c}{Per-target metrics} \\
			\cmidrule{4-7}\cmidrule{9-12}
			Method & z-score & & MSE & $R^2$ & Pearson, $r$ & Spearman, $\rho$& & MSE & $R^2$ & Pearson, $r$ & Spearman, $\rho$\\
			\midrule
			SBROD\citep{karasikov2019smooth}&$1.282$& &$0.961$&$-81.899$&$0.552$&$0.531$& &$0.961$&$-427.838$&$0.762$&$0.685$\\
			VoroMQA\citep{olechnovivc2017voromqa}&$1.410$& &$0.051$&$-3.426$&$0.675$&$0.700$& &$0.051$&$-19.762$&$0.803$&$0.766$\\
			ProQ3\citep{Uziela2016}&$1.670$& &$0.035$&$-2.036$&$0.795$&$0.806$& &$0.035$&$-16.572$&$0.801$&$0.750$\\
			Ornate\citep{pages2019protein}&$1.780$& &$0.007$&$0.424$&$0.813$&$0.805$& &$0.007$&$-1.101$&$\boldsymbol{0.828}$&$\boldsymbol{0.781}$\\
			VoroCNN\citep{igashov2020vorocnn}&$\boldsymbol{1.871}$& &$0.007$ & $0.370$&$0.818$&$0.803$&&$0.007$&$-1.380$&$0.817$&$0.774$\\
			& & & & & & & & & & \\
			Baseline&$1.025$& &$0.011$&$0.065$&$0.658$&$0.666$& &$0.011$&$-2.641$&$0.677$&$0.604$\\
			S-GCN(5)&$1.704$& &$0.010$&$0.157$&$0.854$&$0.831$& &$0.010$&$-1.890$&$0.797$&$0.738$\\
 			S-GCN(10)&$1.665$& &$\boldsymbol{0.005}$&$\boldsymbol{0.573}$&$0.812$&$0.789$& &$\boldsymbol{0.005}$&$\boldsymbol{-0.831}$&$0.710$&$0.680$\\
			S-GCN$_{\text{s}}$(5)&$1.609$& &$0.015$&$-0.272$&$\boldsymbol{0.872}$&$\boldsymbol{0.853}$& &$0.015$&$-3.866$&$0.816$&$0.762$\\
			S-GCN$_{\text{s}}$(10)&$1.303$& &$0.006$&$0.492$&$0.803$&$0.790$& &$0.006$&$-0.917$&$0.738$&$0.683$\\
			\bottomrule
		\end{tabular}
	}
	\label{table:casp12}
\end{table}

\begin{table}[t!]
	\centering
	\caption{
	Comparison of S-GCN and S-GCN$_{\text{s}}$ with the baseline network and the state-of-the-art MQA methods on the unrefined CASP13 stage2 dataset. 
	Parameters in parentheses correspond to the order of the spherical harmonic expansion.}
		\resizebox{\textwidth}{!}{\begin{tabular}{@{}llcllllcllll@{}}
			\toprule
			& 			            & & \multicolumn{4}{c}{Global metrics} & & \multicolumn{4}{c}{Per-target metrics} \\
			\cmidrule{4-7}\cmidrule{9-12}
			Method & z-score & & MSE & $R^2$ & Pearson, $r$ & Spearman, $\rho$& & MSE & $R^2$ & Pearson, $r$ & Spearman, $\rho$\\
			\midrule
			SBROD\citep{karasikov2019smooth}&$1.453$& &$0.050$&$-3.234$&$0.417$&$0.433$& &$0.051$&$-22.455$&$0.805$&$0.761$\\
			VoroMQA\citep{olechnovivc2017voromqa}&$1.369$& &$0.038$&$-2.197$&$0.659$&$0.688$& &$0.038$&$-15.930$&$0.804$&$0.768$\\
			ProQ3\citep{Uziela2016}&$1.459$& &$0.035$&$-1.969$&$0.726$&$0.728$& &$0.035$&$-17.519$&$0.775$&$0.737$\\
			Ornate\citep{pages2019protein}&$1.403$& &$0.009$&$0.193$&$0.786$&$0.799$& &$0.009$&$-2.326$&$0.814$&$\boldsymbol{0.786}$\\
			VoroCNN\citep{igashov2020vorocnn}&$1.516$& &$\boldsymbol{0.007}$&$0.368$&$0.764$&$0.767$& &$\boldsymbol{0.007}$&$-1.962$&$0.811$&$0.771$\\
			&&&&&&&&&&& \\
			Baseline&$0.865$& &$0.017$&$-0.424$&$0.465$&$0.491$& &$0.017$&$-6.375$&$0.648$&$0.619$\\
			S-GCN(5)&$1.362$& &$0.013$&$-0.118$&$\boldsymbol{0.806}$&$\boldsymbol{0.808}$& &$0.013$&$-3.459$&$0.789$&$0.744$\\
			S-GCN(10)&$1.247$& &$\boldsymbol{0.007}$&$\boldsymbol{0.422}$&$0.774$&$0.783$& &$\boldsymbol{0.007}$&$\boldsymbol{-1.348}$&$0.722$&$0.694$\\
			S-GCN$_{\text{s}}$(5)&$\boldsymbol{1.582}$& &$0.020$&$-0.668$&$0.801$&$0.799$& &$0.020$&$-6.415$&$\boldsymbol{0.820}$&$0.773$\\
			S-GCN$_{\text{s}}$(10)&$1.281$& &$0.008$&$0.336$&$0.779$&$0.785$& &$0.008$&$-1.760$&$0.742$&$0.702$\\
			\bottomrule
		\end{tabular}
	}
	\label{table:casp13}
\end{table}

\subsection{Results}

We compared S-GCN and S-GCN$_{\text{s}}$ with our baseline network architecture, and also with the state-of-the-art
single-model \citep{cheng2019estimation}
quality assessment methods SBROD \citep{karasikov2019smooth}, VoroMQA \citep{olechnovivc2017voromqa},  ProQ3 \citep{Uziela2016}, Ornate \citep{pages2019protein}, and VoroCNN \citep{igashov2020vorocnn}.
SBROD is a regression-based method operating on 4D geometric descriptors,
VoroMQA uses statistics from Voronoi 3D tessellation,
ProQ3 is a neural-network-based method with precomputed descriptors of various origin,
Ornate uses deep convolutional networks to process volumetric data in local coordinate frames, and, finally,
VoroCNN is a graph convolutional network built on an atom-level molecular graph.
We downloaded the results of VoroMQA and ProQ3 for CASP[12-13] and the results of SBROD for CASP13 from the official CASP archive at \url{predictioncenter.org}. 
To obtain the results of SBROD on CASP12 and Ornate and VoroCNN on CASP[12-13], we ran these methods locally.
We should emphasize that the main results we report in this work were obtained on the CASP13 dataset, which was not used during training and validation. 
However, to give a complete picture, we also provide the results obtained on the unrefined CASP12 dataset. 
Table \ref{table:casp12} lists the results for CASP12 and Table \ref{table:casp13} lists the results for CASP13.

\begin{figure}[t!]
	\centering
	\includegraphics[width=\textwidth]{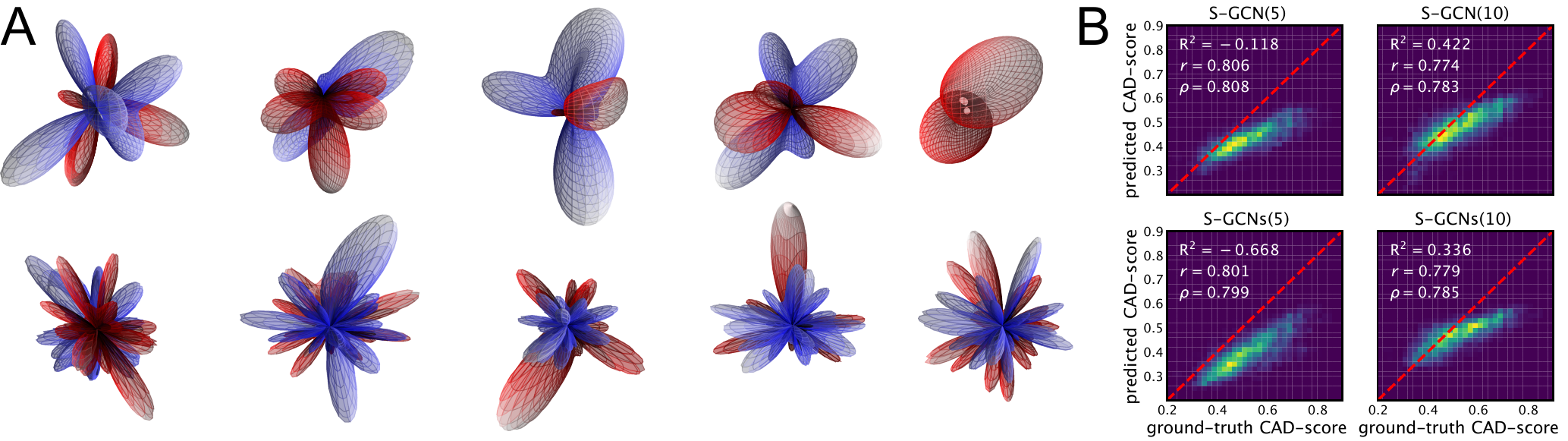}\\
	\caption{
	\label{fig:filters} 
	(A) Examples of spherical filters learned by S-GCN of order $5$ (top row) and S-GCN of order $10$ (bottom row). 
	The distance to the center is proportional to the absolute function value. The red color corresponds to the positive values, the blue color -- to the negative ones.
	(B) Histograms comparing the ground-truth scores and the predictions of spherical graph convolutional networks on the CASP13 dataset.}
\end{figure}

First of all,  we can see a huge performance gap between the baseline network and the other methods.
This can be explained by the fact that the baseline approach uses neither the 3D structure of the graph nor additional chemo-physical or biological features that are widely accepted by the state-of-the-art methods.
At the same time, we would like to emphasize that our spherical graph convolutional networks, which explicitly use the 3D structure of the data, managed to achieve a similar or better quality of predictions compared to the state-of-the-art methods.
Figure \ref{fig:filters}A shows some of the spherical filters learned by the 5th and the 10th order S-GCNs.
We can see their rather complex shape, which is difficult to interpret solely from physico-chemical considerations.

Tables  \ref{table:casp12} and  \ref{table:casp13} also demonstrate
that using a higher order of the spherical harmonic expansion improves the
MSE and $R^2$ metrics. 
At the same time, the order 5 seems to outperform the 10th order in correlation and z-score metrics.
This behaviour becomes clearer if we look closer at the absolute values of the predictions.
Indeed, Figure \ref{fig:filters}B illustrates that the predictions of the 10th order network are closer to the diagonal, thus improving MSE and $R^2$.
It also explains that even though some methods can have high correlation metrics, their predictions are shifted with respect to the main diagonal, 
which results in negative $R^2$. 
Thus, we can conclude that a higher polynomial order of the network allows us to better predict the absolute values of protein scores.

Regarding the  correlation metrics, the 5th order S-GCN performs better than the others.
We can conclude that it should be the method of choice
for ranking protein models and selecting the best model from a given set.
Also, taking into account the fact that the 10th order S-GCN has considerably more trainable parameters, 
and takes 4 times more disk space than the 5th order S-GCN,
it makes more sense to use the latter for practical tasks. 
We can also see
that the last scoring layer improves the correlation metrics.
S-GCNs also demonstrate a  better prediction quality on z-scores, but, as we show in Supplementary Materials, 
these metrics are not stable and we can not confidently say that one method outperforms another because they all have intersecting confident
intervals.

One final remark that we can make after comparing the 5th and the 10th order  S-GCNs is that the 10th order architecture may require significantly more training data.
Therefore, the current CASP training set may not be very well suited for higher-order architectures.
As an alternative, one can consider training on Rosetta-generated decoys \citep{hiranuma2020improved} or using other methods for protein structure prediction.


\section{Conclusion}
In this work,  we applied spherical convolutions to capture the 3D structure of a protein graph. 
The results demonstrate that  our method
gives a significant improvement in the quality of predictions compared to the  baseline without orientational relations between the graph nodes.  
The spherical convolution method can also be
combined with other approaches for the protein model quality assessment, and can also potentially use more input features. 
Thus, we believe it will be possible to achieve even higher prediction results adding biological and chemical information to the input graphs. 
In addition, we would like to notice that the idea of spherical convolutions is universal and can be applied to
various types of  graph-learning tasks, provided that 
the graph structure permits us to define an equivariant coordinate systems for each graph node.

\section{Acknowledgements}
We would like to thank Kliment Olechnovic from Vilnius University for his help on graph construction and active discussions during the project.
We would also like to thank Elodie Laine from Sorbonne Université for the discussions during the study and proof-reading the manuscript,
and Jinbo Xu from Toyota Technological Institute at Chicago for his helpful comments on the manuscript.
This work was partially supported  by the Inria International Partnership program BIOTOOLS.

\section*{ Broader Impact }

Our work will likely stimulate the development of new representation-learning methods applied to 3D graphs. The latter may represent molecules, such as proteins or nucleic acids. However, these graphs can also describe more general 3D data from other research domains, e.g., from astronomy or earth science. 

From the application point of view, we believe our method will be useful in the structural bioinformatics and structural biology communities. 
Indeed, there has been a very rapid improvement of methods for 3D protein structure prediction, mostly owing to novel developments in deep learning and in algorithms extracting coevolution signals from sequence data. 
However, the question of how to assess the quality of the predicted models, and which parts of the predicted structures are likely to be less accurate, is still open. Thus, we hope that the proposed approach will help bioinformaticians and structural biologists to better use and analyze available computational data. 

From a more general perspective, proteins are responsible for the main cellular functions in any organism. 
They maintain the shape of the cells, control chemical catalysis, play the role of cellular motors, and regulate vital processes. 
This makes the study of protein structures and interactions an important part of molecular biology. 
Understanding protein structure is also crucial for therapeutic purposes toward the development of new drugs, and the improvement of existing ones. 
To conclude, gaining knowledge of proteins and their functions contribute to a better understanding of the life machinery and organization, which has a significant social impact. 

\bibliographystyle{plainnat}
\bibliography{Pavlichenko2020Project52}

\end{document}